\documentclass[aps,prb,twocolumn,superscriptaddress,amsmath,showpacs]{revtex4-1}

\usepackage[pdftex]{graphicx}%
\usepackage{bm}

\begin{document}

\title{Orbital upper critical field of type-II superconductors with pair breaking }

\author{V. G. Kogan}
\email{kogan@ameslab.gov}
\affiliation{The Ames Laboratory, Ames, IA 50011, U.S.A.}

\author{R. Prozorov}
\email{prozorov@ameslab.gov}
\affiliation{The Ames Laboratory, Ames, IA 50011, U.S.A.}
\affiliation{Department of Physics \& Astronomy, Iowa State University, IA 50011, USA}

\begin{abstract}
 The orbital upper critical field   $H_{c2} $  is evaluated  for isotropic materials with arbitrary  transport and pair-breaking scattering rates. It is shown that unlike transport scattering which enhances $H_{c2} $, the pair breaking  suppresses the upper critical field and reduces the dimensionless ratio $h^*(0)=H_{c2}(0)/T_c(dH_{c2}/dT)_{T_c} $ from the Helfand-Werthamer value of $\approx 0.7$ to 0.5 for a strong pair-breaking. $h^*(T)$ is evaluated for arbitrary  transport and pair-breaking scattering.  
 A phenomenological model for   the pair-breaking suppression by  magnetic fields is introduced. It shows  qualitative features such as a positive curvature of $H_{c2}(T) $   and the low temperature upturn  usually associated with  multi-band superconductivity. 
\end{abstract}



\date{25 May, 2013}  

\maketitle

\section{Introduction}

In a seminal work Helfand and Werthamer calculated the $H_{c2}(T)$
 for isotropic materials with non-magnetic impurities.\cite{HW} In particular, they showed that the ratio $H_{c2}(0)/T_c H_{c2}^\prime (T_c)\approx 0.7$ for any impurity
content.  Since then, this result is broadly used to estimate $H_{c2}(0)$ by measuring
a readily accessible slope $H_{c2}^\prime$ at $T_c$ although many new materials of interest are anisotropic with a substantial    pair-breaking scattering. 
 
 The general $H_{c2}(T)$ problem for materials with anisotropic Fermi surfaces and  order parameters is quite complicated\cite{MMK,Gur1,Gur2,KP-ROPP,Kita} and applying the existing models to real materials requires knowledge of many material parameters. Analyzing the $H_{c2}$ data, conclusions are often made just on the basis of analogy with other materials. An example is a commonly held belief that a positive curvature of the $H_{c2}$ curve near $T_c$ is an evidence for a multi-gap scenario analogous to the well studied MgB$_2$.
 
 In this work we have a less ambitious goal of solving the one-band isotropic problem in the presence of both transport and pair breaking     scattering. This problem has been considered by Fulde and Maki in a more general context of correlated magnetic impurities.\cite{FM} They, however, considered only the limit of short transport scattering time. On the other hand,   clean materials with a strong pair breaking can  in principle  exist, CeCoIn$_5$ is an example.\cite{KPPetr}
 
 We take advantage of numerical methods now available and show that various combinations of scattering rates, $1/\tau$ and   $1/\tau_m$ ($ \tau_m$ is the pair-breaking, e.g., spin-flip,  scattering time) may cause variety of behaviors of $H_{c2}(T)$ which might be useful interpreting the data on real materials at least on a qualitative level. 
 
 In the second, more speculative, part of this work we discuss an interesting possibility: The rate $1/\tau_m$ of the spin-flip scattering of
conducting carriers on local moments may depend on the applied 
field because the  spin flip should be accompanied by a
change of the spin associated with local moments, the
energy of the latter is $H$ dependent. We have included this possibility within our formalism and obtained variety of behaviors of $H_{c2}(T)$ which open yet another channel in interpretation of the temperature dependence of the upper critical field.

\section{The problem of $H_{c2}$}

 Consider an isotropic material with both magnetic and
non-magnetic scatterers. The problem of the 2nd order
phase transition at 
$H_{c2}$  is addressed using the Eilenberger quasiclassical
version\cite{Eil} of Gor'kov's equations for normal and anomalous Green's
functions $g$ and $f$. At $H_{c2}$, $g=1$ and we are left with a linear equation for $f$:
\begin{eqnarray}
(2\omega^+ +~\textbf{v}\cdot {\bm \Pi})\,f=2\Delta
/\hbar+ \langle f\rangle/\tau^- \,,\label{eil1}\\
\omega^+=\omega +\frac{1}{2\tau^+}\,,\quad \frac{1}{\tau^\pm}=\frac{1}{\tau }\pm
\frac{1}{\tau_m}\,.
\end{eqnarray}
Here, ${\bm v}$ is the Fermi velocity, ${\bm  \Pi} =\nabla +2\pi i{\bm 
A}/\phi_0$ with    the vector potential $\bm A$ and  the flux quantum $\phi_0$. $\Delta ({\bm  r})$ is the gap
function (the order parameter); 
the  Matsubara frequencies are defined by $\hbar\omega=\pi T(2n+1)$
with an integer $n$; $\langle...\rangle$ stand for  averages over
the Fermi surface.  

Solutions $f$ and $\Delta$ of Eq.\,(\ref{eil1}) should satisfy the  self-consistency equation:
\begin{equation}
\frac{\Delta}{2\pi  T}\ln\frac{T_{c0}}{T}=\sum_{\omega>0}\left(\frac{\Delta}
{\hbar\omega }-\langle f\rangle\right)\,,
\label{selfcons0}
\end{equation}
where $T_{c0}$ is the critical temperature in the absence of pair-breaking scattering. 
In zero field, Eq.\,(\ref{eil1}) yields 
\begin{equation}
\langle f\rangle= \frac{\Delta} {\hbar \omega_m}\,,\qquad \omega_m = \omega +\frac{1}{\tau_m} \,.
\label{omega'}
\end{equation}
Substituting this in Eq.\,(\ref{selfcons0}) one obtains an equation for the actual $T_c$ which together with  Eq.\,(\ref{selfcons0}) allows one to exclude $T_{c0}$:
\begin{equation}
\frac{\Delta}{2\pi
T}\ln\frac{T_c}{T}=\sum_{\omega>0}\left(\frac{\Delta}
{\hbar\omega^\prime}-\langle f\rangle\right),\quad \omega^\prime=\omega+\frac{t}{\tau_m} 
\label{selfcons}
\end{equation}
where $T_c$ is the actual (suppressed by magnetic impurities) critical
temperature  and $t=T/T_c$.

The general scheme for finding $H_{c2}(T;\tau,\tau_m)$ is as follows:
The solution of Eq.\,(\ref{eil1}) is written in the form:
\begin{equation}
f={2\over \hbar}\int_0^{\infty}d\eta\,e^{-\eta (2\omega^+ +~\textbf{v}\cdot
{\bm\Pi})} \left(\Delta+\frac{\hbar \langle f\rangle}{2\tau^-}\right)\,.
\label{eq4}
\end{equation}
Taking average over the Fermi surface of both sides we have:
\begin{equation}
F={2\over \hbar}\int_0^{\infty}d\eta\,e^{-2\eta  \omega^+}\Big\langle
e^{-\eta~\textbf{v}\cdot {\bm\Pi}}\Big\rangle \left(\Delta+\frac{\hbar
F}{2\tau^-}\right)\,,
\label{F}
\end{equation}
where $F=\langle f\rangle$. As argued in Refs.\,\onlinecite{HW,coherence} 
both $\Delta$ and $F$ satisfy at  $H_{c2}(T)$ a linear
equation  $- \xi^2 \Pi^2 \Delta =\Delta$ which gives $H_{c2}=\phi_0/2\pi\xi^2$.  This allows one to manipulate
the exponential operator to the form
\begin{equation}
e^{-\eta~\textbf{v}\cdot {\bm\Pi}} \Delta
=\Delta\,\exp\left(-\frac{\eta^2v_\perp^2 }{4\xi^2}\right)\,,
\label{exp}
\end{equation}
and the same for $F$; $v_\perp$ is the Fermi velocity projection onto the plane perpendicular to $\bm H$. The Fermi sphere average of this expression is
readily found:
\begin{equation}
      \Big\langle
e^{-\alpha^2\sin^2\theta}\Big\rangle =
\frac{\sqrt{\pi}}{2\alpha}\,e^{-\alpha^2} {\rm Erfi}(\alpha)\,,\quad
\alpha=\frac{\eta v  }{2\xi }\,,
\label{average}
\end{equation}
where $\theta$ is   the polar angle on the sphere, 
${\rm Erfi}(\alpha)={\rm erf}(i\alpha)/i=(2/\sqrt{\pi})\int_0^\alpha
dt\,e^{t^2}$.  Substituting this in (\ref{F}) we find $F(\bm r)\propto \Delta(\bm r)$:
\begin{eqnarray}
F &=& \frac{2\tau^-\Delta}{\hbar}\,\frac{J}{ \tau^--J} \,,\label{F_Delta}\\
J(\xi,T,\tau^+)&=&\frac{\sqrt{\pi}}{2 }\,\int_0^\infty d\eta
\,e^{-2  \omega^+\eta} \,\frac{e^{-\alpha^2}}{\alpha}\,
{\rm Erfi}(\alpha).\qquad
\label{J}
\end{eqnarray}
Hence, we have the self-consistency relation:
\begin{equation}
\frac{1}{2\pi
T}\ln\frac{T_c}{T}=\sum_{\omega>0}\left(\frac{1}
{\hbar\omega^\prime}-\frac{2\tau^- J}{\hbar(
\tau^--J)}\right)\,,
\label{selfcons4}
\end{equation}
which is an equation for $\xi(T;\tau,\tau_m)$.
It is readily seen that this  equation  reduces to
the standard form  for non-magnetic scattering if one sets $\tau_m\to\infty$.
 
The integral $J$ is convergent; this is seen from the power series\cite{Abr}
\begin{equation}
\frac{{\rm Erfi(\alpha)}}{ \alpha}\,e^{-\alpha^2}
=\frac{2}{\sqrt{\pi}} \sum_{n=0}^\infty \frac{(-2)^n\alpha^{2n}}{(2n+1)!!}\,,
\label{series}
\end{equation}
which gives a constant for $\alpha\to 0$. We can use this expansion to
recast $J$ in a different form. To this end, substitute it in
Eq.\,(\ref{J}) and integrate:
\begin{equation}
J = \frac{1}{ 2\omega^+} \sum_{n=0}^\infty
\frac{(-1)^nn! }{ 2n+1 }\left(\frac{v}{2\xi\omega^+}\right)^{2n}\,.
\label{J1}
\end{equation}
The sum here belong to   Borel summable
types.\cite{Hardy} It has been studued by HW and can be written as an
integral
\begin{equation}
J = \frac{2\xi}{v}\,\int_0^\infty du
\,e^{-u^2} \tan^{-1}\left(\frac{v}{2\xi\omega^+}\,u\right).
\label{J2}
\end{equation}
Another integral representation is given in Ref.\,\onlinecite{KN}:
\begin{equation}
J = \frac{\sqrt{\pi}\,\xi}{v}\,\int_0^\infty \frac{dt}{1+t^2}
\,{\rm erfc}\left(\frac{2\xi\omega^+}{v}\,t\right).
\label{J3}
\end{equation}


We now introduce  dimensionless variables
\begin{equation}
t = \frac{T}{T_c}\,,\qquad h=
  \frac{\hbar^2v^2}{ 4\pi^2T_c^2\xi^2 } =H_{c2}\frac{\hbar^2v^2}{
2\pi T_c^2 \phi_0} \,,
\label{t,h}
\end{equation}
and the scattering parameters
\begin{equation}
\rho_m = \frac{\hbar}{2\pi T_c\tau_m}\,,\quad \rho  =
\frac{\hbar}{2\pi T_c\tau }\,,\quad \rho^\pm =\rho\pm\rho_m\,.
\label{rhos}
\end{equation}
Note that $\rho,\rho_m$ involve the actual $T_c$, they  differ  from   used often scattering parameters defined via $T_{c0}$.
 
The self-consistency  Eq.\,(\ref{selfcons4}) in dimensionless form reads:
\begin{eqnarray}
&&    -\ln t=  \sum_{n=0}^\infty\left(\frac{1
}{ n+1/2+\rho_m }-\frac{2tI}{1-\rho^-I }\right)\,,\label{eq20}\\
&&I=\sqrt{\pi}\gamma\int_0^\infty\frac{dz\,{\rm erfc}\,z}
{z^2h+\gamma^2}\,,\quad \gamma = t(2n+1)+\rho^+\,.\qquad
\label{Igamma}
\end{eqnarray}
This can be solved numerically for $h(t)$ for any combination of
scattering parameters $\rho $ and $\rho_m$.

\subsection{  ${\bm T}\to {\bm T_c} $}

As $T\to T_c$, $h\to 0$ and the parameter 
\begin{equation}
 s  =   \sqrt{h}\, /\gamma \,,
\label{eta}
\end{equation}
can be considered as small. The integral $I$ can then be evaluated:
\begin{eqnarray}
I&=&\frac{\sqrt{\pi}}{\gamma}\int_0^\infty\frac{dz\,{\rm erfc}\,z}
{z^2s^2+1} \nonumber\\ 
&\approx& \frac{\sqrt{\pi}}{\gamma}\int_0^\infty dz\, (1-z^2s^2){\rm
erfc}\,z 
= \frac{1}{\gamma}\left(1-\frac{s^2}{3}\right);\qquad    
\label{I2}
\end{eqnarray}
 ${\rm erfc}(z)$ effectively truncates the integration domain to
approximately $z<2$, so that the expansion of $(1+z^2s^2 )^{-1}$ in powers of $z^2s^2$ is justified. 
We then obtain keeping only the terms $\sim s^2$ in Eq.\,(\ref{eq20}): 
\begin{eqnarray}
    -\ln t&=& \psi\left( \frac{ \rho_m }{  t}+\frac{1}{2}\right) -\psi\left( 
\rho_m+\frac{1}{2}\right)\nonumber\\
 &+&\frac{h}{3 \rho_-^2}\Big[
\psi\left(\frac{ \rho_m }{ t}+\frac{1}{2}\right)
-\psi\left(\frac{ \rho_+ }{2t}+\frac{1}{2}\right)\nonumber\\
&+&
\frac{\rho_-}{2t}\psi^\prime \left(\frac{
\rho_m}{t}+\frac{1}{2}\right)
\Big].\label{s-c_small_eta} 
\end{eqnarray}
Expanding this in powers of $1-t\ll1$, we obtain the slope at $t=1$:
\begin{eqnarray}
 &&    -\frac{dh}{dt}\Big|_{t=1}=3 \rho_- ^2\left[1-\rho_m\psi^\prime \left( 
\rho_m +\frac{1}{2}\right)\right]  \Big/\qquad \qquad \qquad \nonumber\\
&&\left[ \psi\left(\rho_m +\frac{1}{2}\right) 
- \psi\left(\frac{\rho^++1}{2}\right) + \frac{\rho_-}{2} \psi^\prime \left( 
\rho_m +\frac{1}{2}\right)\right]  .\qquad\label{h'_GL}
\end{eqnarray}
 If $\rho_m=0$, this reduces to the HW result for  non-magnetic scattering.\cite{HW}
 \begin{figure}[htb]
 \includegraphics[width=9cm]{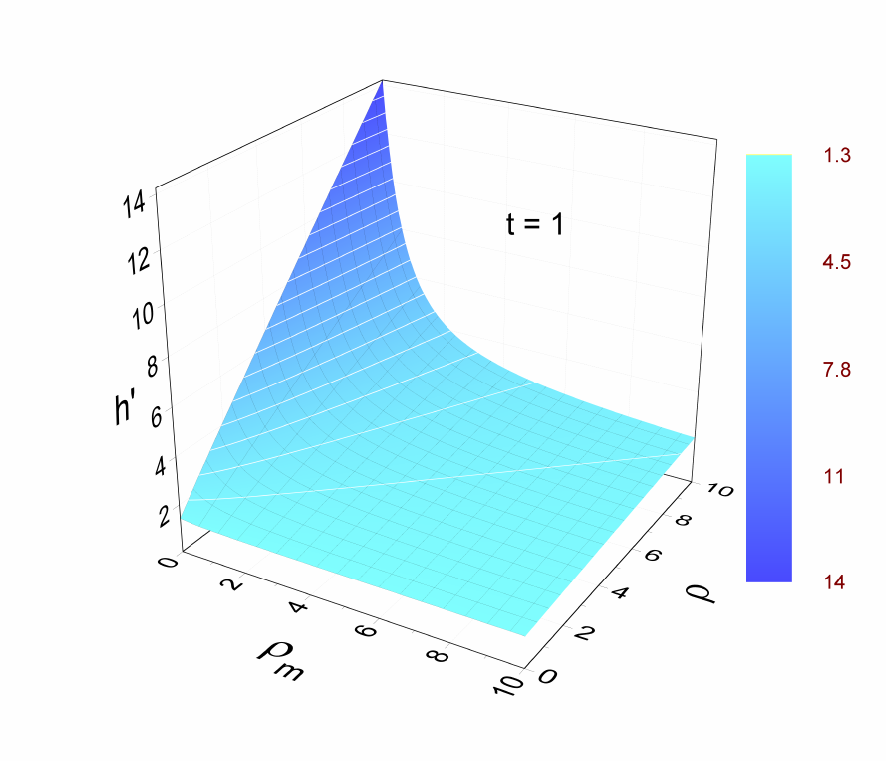}
 \caption{(Color online)  The slope  $| dh/dt | $ at $ t=1 $ as a function of two scattering parameters: $0<\rho<10$, $0<\rho_m<10$. Withe lines show contours of $h^\prime$=const.}
  \label{slope}
 \end{figure}

 Fig.\,\ref{slope} shows the slopes according to Eq,\,(\ref{h'_GL}). One observes that the pair-breaking scattering depresses the slopes $h^\prime$ at $T_c$, just the opposite to what the transport scattering does. We see that (i) for  weak transport scattering $\rho$, the slopes are nearly independent of the magnetic scattering $\rho_m$, and (ii) for strong pair-breaking scattering (roughly $\rho_m >4$)  the slopes remain low even if the transport scattering intensifies. 

In common units, the slope
\begin{equation}
\frac{dH_{c2}}{dT}\Big|_{T_c}=  \frac{2\pi  \phi_0}  {\hbar^2v^2}\,T_c\frac{dh}{dt}\Big|_{t=1}.
\end{equation}
Hence, one can say that in a broad domain of scattering parameters 
\begin{equation}
\frac{dH_{c2}}{dT}\Big|_{T_c}\propto T_c  
\end{equation}
provided roughly $\rho_m >4$. This feature, in fact, has been suggested as evidence of a  pair  
breaking present in many iron-based superconductors.\cite{Kog1,Kog2,Kog3}

\subsection{ Strong pair breaking, ${\bm T_c\to 0}$}

When   $\tau_m$ is close to the critical value where $T_c\to
0$, $H_{c2}$ can be calculated analytically in the whole temperature
range $0<T<T_c$. Formally, the simplification comes about because in this
domain all $\rho$'s are large. 
Then,   $ s  =   \sqrt{h}\, /\gamma$   is small due to large $\gamma$.   Eq.\,(\ref{I2}) and (\ref{s-c_small_eta}) are still valid and one can do sums in Eq.\,(\ref{eq20}) keeping only terms ${\cal O}(s^2)$. We can utilize the asymptotic expansion $\psi(x+1/2)=\ln
x+1/24x^2+{\cal O}(1/x^4)$ to obtain:
\begin{equation}
h= \frac{1}{8}\left(\frac{\rho^-
}{\rho_m }\right)^2\left(\frac{\rho^-}{2\rho_m}+\ln
\frac{2\rho_m}{ \rho^+}\right)^{-1} (1-t^2)\,.
\label{h_gapless}
\end{equation}
It is worth noting that here the ratio
\begin{eqnarray}
 h^*(0)=  \frac{ H_{c2}(0)}{T_c|H_{c2}^\prime(T_c)|}=\frac{h(0)}
{ h^\prime (1)} = \frac{1}{ 2}   \,. 
\label{ratio} 
\end{eqnarray}
 The value $h(0; \rho,\rho_m)$ as given in  Eq.\,(\ref{h_gapless}) in fact depends only on the ratio $\rho/\rho_m$ and varies
from the minimum  of $1/4\ln(4/e)=0.647$ corresponding to $\rho/\rho_m\ll 1$, through the 
unity at $\rho/\rho_m= 1$, to $\rho/4\rho_m $ for $\rho/\rho_m\gg 1$.
For the gapless regime with $\rho\gg
\rho_m$,   Eq.\,(\ref{h_gapless}) reduces to the result   of Abrikosov and Gor'kov.  \cite{AG}

\subsection{Numerical results }

Equations (\ref{eq20}) and (\ref{Igamma}) can be solved numerically for any $\rho$ and $\rho_m$. Numerical results were obtained using Matlab and Mathematica. Attention has to be paid to the number of summation terms in Eq.\,(\ref{eq20}). At low temperatures as many as 5000 terms were needed. 

\begin{figure}[htb]
 \includegraphics[width=8cm]{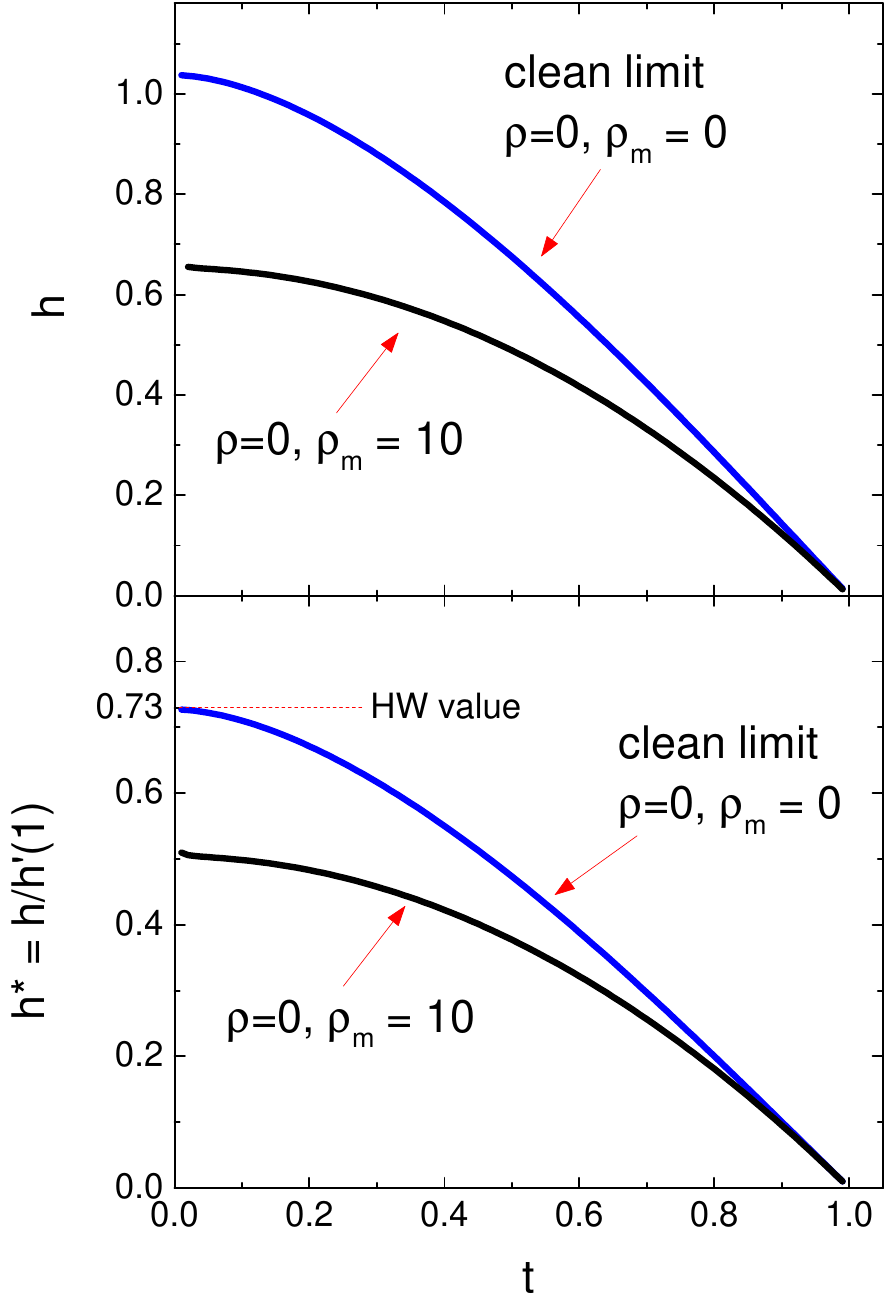}
 \caption{(Color online)  The upper panel: the clean limit $h(t)$; the lower curve  shows that the pair-breaking scattering not only suppresses $T_c$ but suppresses the dimensionless $h(t)$ as well. The lower panel shows the same results in terms of the HW variable $h^*(t)=H_{c2}(T)/T_c(dH_{c2}/dT)_{T_c}=h(t)/h^\prime(1)$. $\rho_m=10$ corresponds to a gapless state with a strong pair-breaking; the numerically obtained value $h(0)/h^\prime(1)=0.5$ is in excellent agreement with Eq.\,(\ref{ratio}).}
  \label{f2}
 \end{figure}

\begin{figure}[htb]
 \includegraphics[width=8cm]{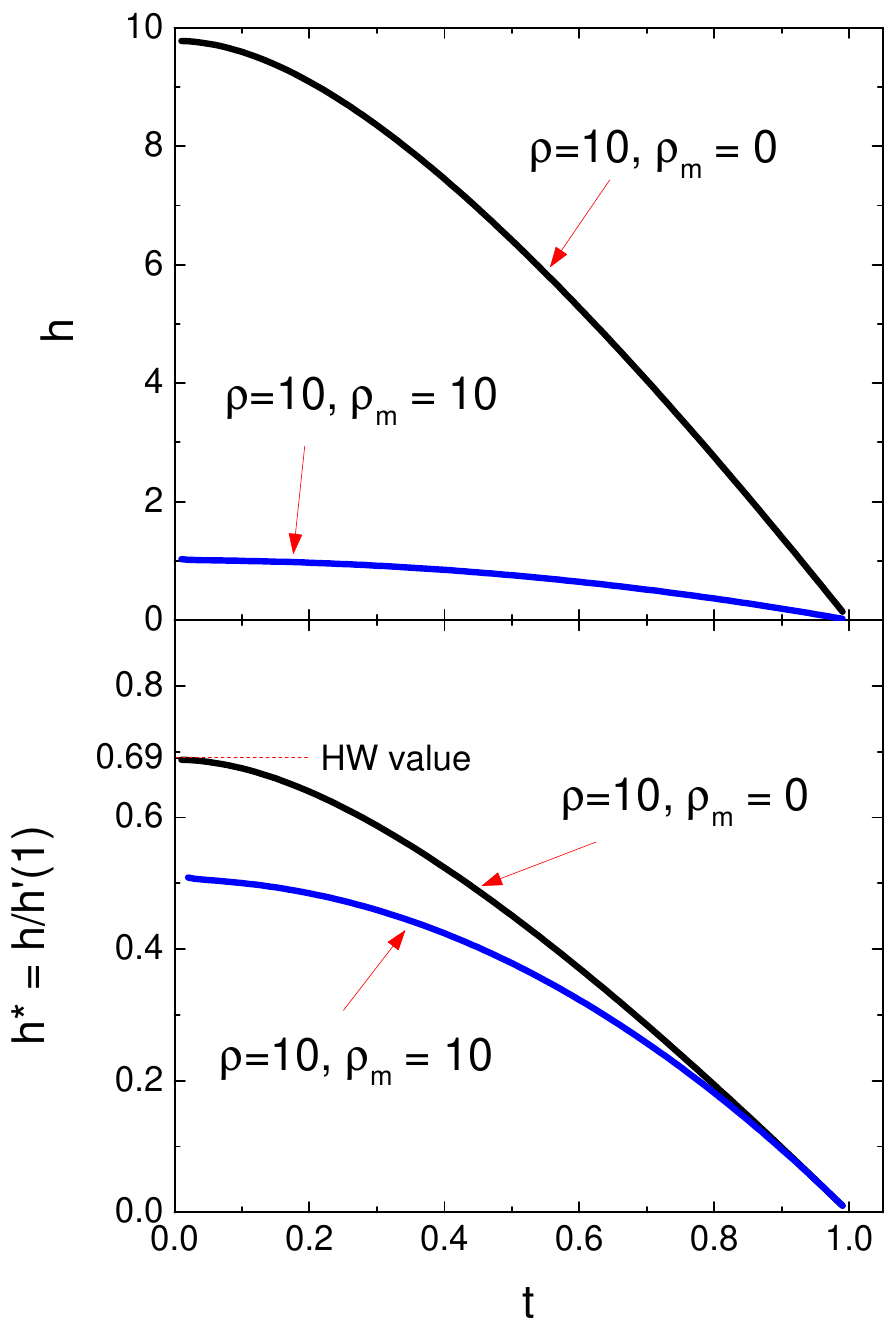}
 \caption{(Color online) The dirty limit curve $h(t)$ calculated for $ \rho=10$ and $ \rho_m=0$ coincides with   HW result which is confirmed by plotting it as the HW reduced variable $h^*(t)=h(t)/h'(1)$ (the lower panel). 
 The   $h(t)$ calculated for $ \rho=\rho_m=10$  in the upper panel shows that unlike transport scattering which enhances the upper critical field, the pair-breaking suppresses $h$.  Comparing $h^*(t)$ of Fig.\,\ref{f2} for $\rho=0,\,\,\rho_m=10$ with $h^*(t)$ of this figure for  $\rho=10,\,\,\rho_m=10$ we conclude that for the strong pair breaking with large $\rho_m$, the transport scattering has practically no effect upon $h^*(t)$.}
  \label{f3}
 \end{figure}

Representative examples of such calculations are given in Fig.\,\ref{f2} and \ref{f3}. Parameters for these graphs are chosen not because they are realistic, but rather to demonstrate evolution of $h(t)$ with changing scattering parameters $\rho$ and $\rho_m$. We also show the HW ratios $h^*(t)=h(t)/h'(1)$ for both clean and dirty transport limits. One clearly sees that this ratio, which is close to 0.7 for purely transport scattering, drops to $\approx 0.5$ for a strong pair-breaking. It is worth noting that actual $H_{c2}(T) $ given in Eq.\,(\ref{t,h}) is $\propto T_c^2$, the latter being suppressed by pair-breaking scattering. Hence, the plots of  $h(t)/h'(1)$ are  valuable in particular. 

 \begin{figure}[htb]
 \includegraphics[width=9cm]{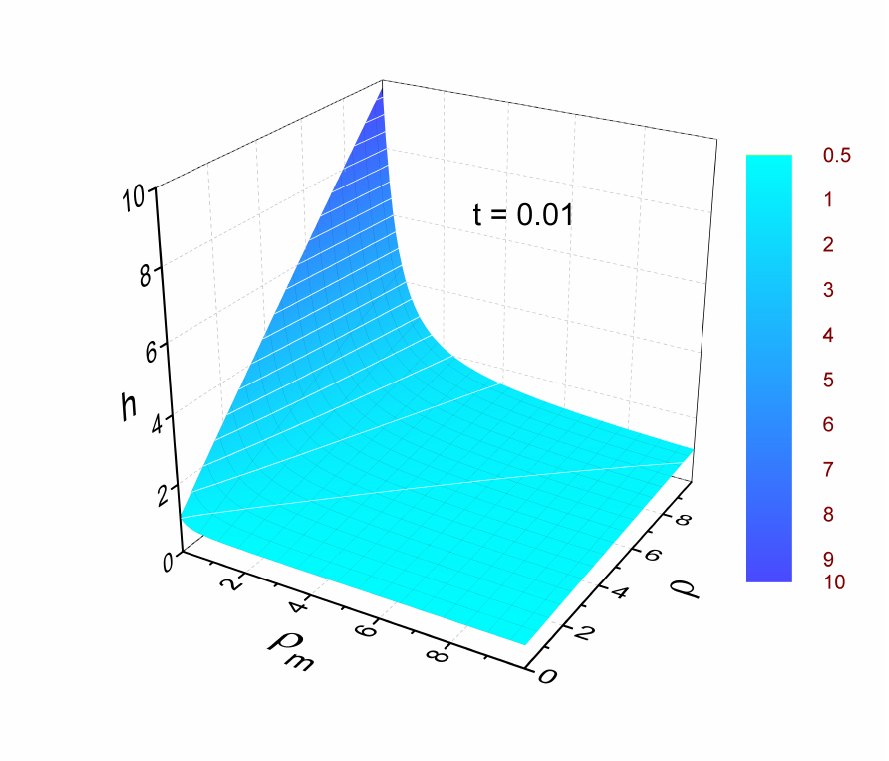}
 \caption{(Color online)  The field  $ h$ at $ t=0.01$  representing $h(0)$  as a function of two scattering parameters: $0<\rho<10$, $0<\rho_m<10$. White lines are contours of $h(0)=\,\,$const.}
  \label{ho}
 \end{figure}
 
 \begin{figure}[h]
 \includegraphics[width=9cm]{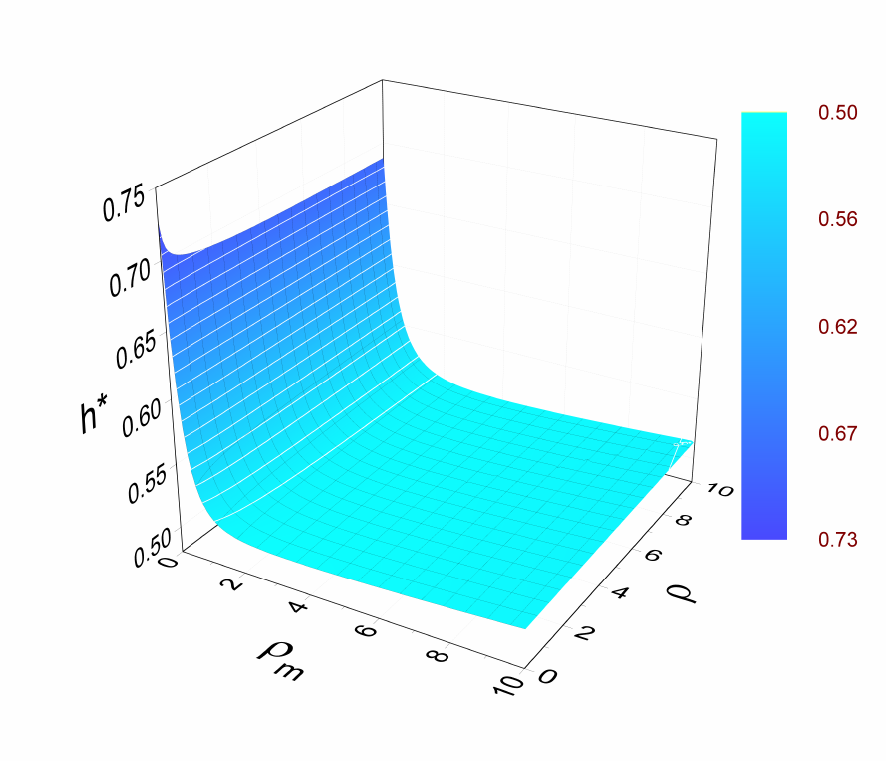}
 \caption{(Color online)  The HW ratio $h^*(0)=h(0)/h^\prime(1)$ {\it versus} scattering parameters  $0<\rho<10$ and $0<\rho_m<10$. White lines are contours of $h^*(0)=\,\,$const.}
  \label{figHW}
 \end{figure}

 Having solved  for $h(t;\rho,\rho_m)$ one can collect the zero-$T$ values  $h(0;\rho,\rho_m)$. This calculation should be done with care because the  sums over $\omega$	 in Eq.\,(\ref{eq20}) are logarithmically divergent and should be truncated at $n$ corresponding to the Debye frequency $\omega_D$: $n_D=\hbar\omega_D/2\pi T$, i.e., it diverges at $t=0$. The calculation then can be done for a small but finite $t$ as shown in Fig.\,\ref{ho}. 
  
 One can now construct the HW ratio $h^*(0)=h(0)/h^\prime(1)$ for any $\rho$ and $\rho_m$ with the result shown in Fig.\,\ref{figHW}. At $\rho_m=0$ we have the standard HW behavior of $h^*(0)$ which is close to 0.73 for the clean limit and reduces to 0.69 at the dirty side. With the pair-breaking increasing,  $h^*(0)$   approaches 0.5 for large $\rho_m$.

\section{Model of field dependent  spin-flip scattering} 
 
The rate $1/\tau_m$ of the spin-flip scattering of
conducting carriers on local moments may depend on the
field because the  spin flip should be accompanied by a
change of the spin associated with local moments, the
energy of the latter is $H$ dependent. 
 If $\delta\mu$ is the local moment change, the probability of  the pair-breaking scattering should contain a factor $\exp(-\delta\mu\, H/T )$. This factor should enter the magnetic scattering parameter: $\rho_m =\rho_{m0} \exp(-\delta\mu H/T )$. Hence, the pair-breaking  scattering becomes weaker with increasing $H$.
	 \begin{figure}[t]
 \includegraphics[width=8cm]{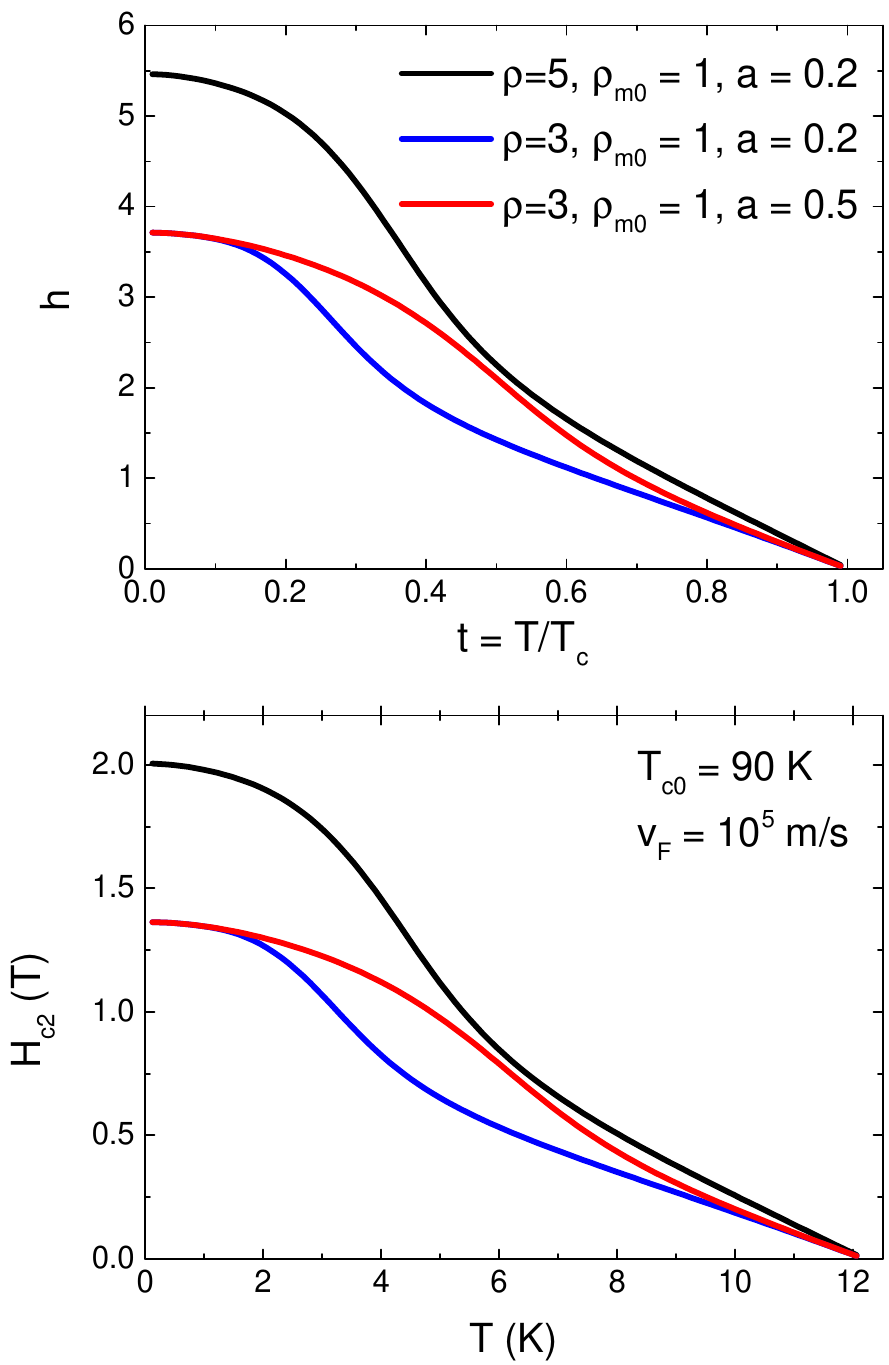}
 \caption{(Color online)  The effect of freezing out of the pair-breaking by field is illustrated on   $h(t)$ calculated for the parameters indicated in the legend. The lower panel shows the same results in common units for a hypothetical material with $T_{c0}=90\,$K and the Fermi velocity $10^5\,$m/s.
}
  \label{f6}
 \end{figure}

  \begin{figure}[h]
 \includegraphics[width=8cm]{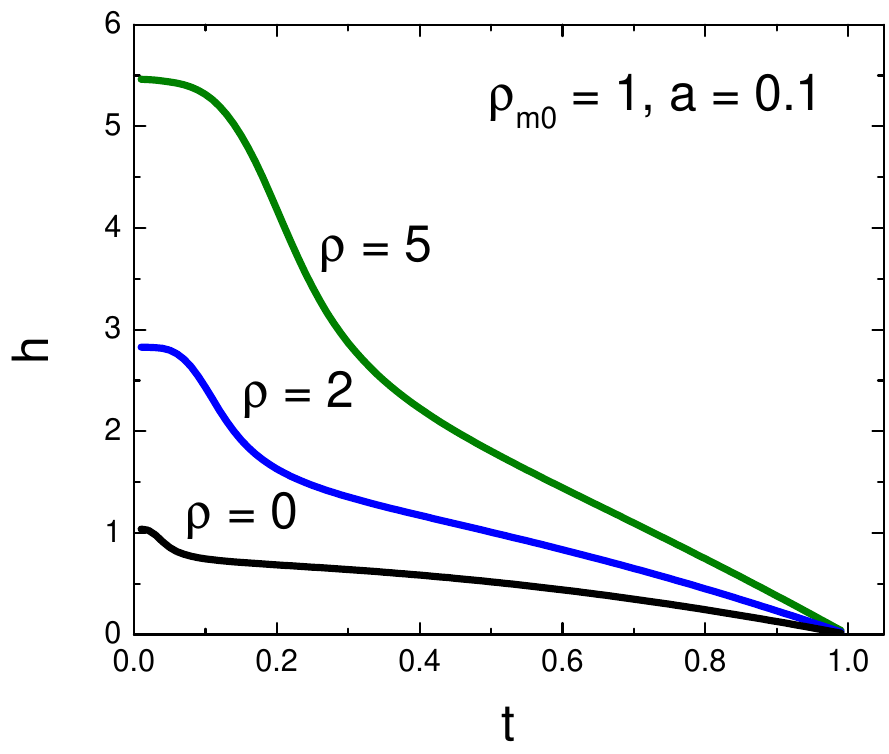}
 \caption{(Color online)   The reduced field $ h(t ) $ for the pair-breaking  scattering parameter  $ \rho_{m0}=1$, the parameter $a=0.1$ and the transport scattering  $\rho= 0, 2 $, and 5. Note the extended near-linear domains of $ h(t ) $ and the upturn at low temperatures, the features seen in a number of Fe-based materials.\cite{Ni-Ni,Gasparov}}
  \label{f7}
 \end{figure}

For an estimate we take $\delta\mu\sim \mu_B$, $\mu_B$ is the Bohr magneton. Then, writing the Boltzmann factor in our dimensionless units as $\exp(-\delta\mu H/T)=\exp(-a h/t)$ one estimates $a\sim 0.03\, T_c(K)$. Setting in our equations for $h(t)$ the parameter 
\begin{eqnarray}
\rho_m=\rho_{m0}\,e^{-a h/t}    
\label{rhoma} 
\end{eqnarray}
we can study qualitatively how the  field suppression of  spin-flip scattering affects $h(t)$. We note that our results for the slopes of $H_{c2}$ at $T_c$ are not affected by this change since there $h\to 0$. On the other hand, as $t\to 0$, the new $\rho_m$ vanishes, i.e., the spin-flip scattering is completely ``frozen out". In the following we will call the constant $a$ the ``pair-breaking freezing parameter". 

 A few examples are given below to illustrate  field effects upon the pair-breaking and their influence on the behavior of $h(t)$. The first interesting feature of the $h(t)$ curve is shown in Fig.\,\ref{f6}: the positive curvature of $h(t)$ at high and intermediate temperatures. Traditionally, this feature is  associated with the multi-band superconductivity, as is the case of MgB$_2$. We now see that the positive curvature of $h(t)$ can be present in a one-band isotropic material due to the pair-breaking scattering and its suppression by the field.

 Figure \ref{f7} shows a set of three curves corresponding to the same magnetic scattering $ \rho_{m0}=1$, the same pair-breaking freezing parameter $a=0.1$, but different transport scattering $\rho=0,2,5$. A   feature of these curves worth noting is nearly linear temperature dependence in a broad temperature domain. This feature is seen in many iron-based materials;\cite{Ni-Ni,Gasparov} our  work therefore suggests that the near-linear behavior of $H_{c2}(T)$ might be related to   pair-breaking.

\section{d-wave} 

We show here that the problem of $H_{c2}$ in a  d-wave material with a spherical Fermi surface in the presence of impurities is simpler than for the s-wave symmetry, because in all relations for   $H_{c2}$  transport and pair-breaking scattering rates enter only via $\rho^+=\rho+\rho_m$. 

Within a popular approximation, the effective coupling  responsible for superconductivity  is assumed factorizable: 
 $ V({\bm  k}_F,{\bm  k}_F^{\prime\,})=V_0 \,\Omega({\bm  k}_F)\,\Omega({\bm  k}_F^{\prime\,})$.\cite{Kad} One  looks for the order parameter in the form  $\Delta (
{\bm  r},T;{\bm  k}_F)=\Psi ({\bm  r},T)\, \Omega({\bm  k}_F)$. The
self-consistency  equation takes the form:
       \begin{equation}
\frac{\Psi }{2\pi T} \ln \frac{T_{c0}}{T}= \sum_{\omega
>0}^{\infty}\left(\frac{\Psi}{\hbar\omega}-\Big\langle \Omega \, f
\Big\rangle
\right)\,.
\label{gap1}
\end{equation}
 $\Omega({\bm  k}_F)$    describes the
variation of $\Delta$ along the Fermi surface and  is  normalized:
$ \langle \Omega^2  \rangle=1$. For the d-wave, $\Omega= \sqrt{2}\cos 2\varphi$ and $ \langle\Delta \rangle=0$.

The Elenberger Eq.\,(\ref{eil1}) holds for any symmetry of the order parameter $\Delta$. 
Taking the average of Eq.\,(\ref{eil1}) in zero field over the Fermi surface we obtain $ \langle   f\rangle =0$ and $f=\Delta/\hbar\omega^+$. Substituting this in Eq.\,(\ref{gap1}) we obtain for the actual critical temperature\cite{Openov,Kog1}
\begin{eqnarray}
   \ln \frac{T_{c0}}{T_c}=   \psi  \left( \frac{\rho^+ +1}{2}\right) -\psi  \left( 
\frac{ 1}{2}\right)
   .\label{Openov}
\label{Tc0/Tc}
\end{eqnarray}
Combining this with Eq.\,(\ref{gap1}) one can exclude $T_{c0}$.

  The same derivation as above results in the  dimensionless form of the self-consistency equation:
\begin{eqnarray}
   -\ln t=  \sum_{n=0}^\infty\left(\frac{1
}{ n+1/2+\rho^+/2 }- 2t\,I \right)\,,\label{s-c-d} 
\end{eqnarray}
This can be solved numerically for $h(t)$ for any  
  $\rho $ and $\rho_m$ which in fact enter only via   $\rho^+=\rho+\rho_m$. 
  
One can obtain slopes $h^\prime(1)$ at the critical temperature in the same manner as for s-wave treatment above:
\begin{eqnarray}
    -\frac{dh}{dt}\Big|_{t=1}=24 \left[1-\frac{\rho^+}{2}\psi^\prime \left( 
\frac{\rho^+ +1}{2}\right)\right]  \Big/ 
  \psi^{\prime\prime} \left( 
\frac{\rho^+ + 1}{2}\right)   .\qquad\label{h'_GLd}
\end{eqnarray}
In the clean limit, this yields $h^\prime = -12/7\zeta(3)$ in agreement with the general clean limit formulas for the d-wave.\cite{KP-ROPP} For a strong $T_c$ suppression when $\rho^+\to \infty$, we get $h^\prime = - 2 $, so that the actual slope at $T_c$ vanishes as $dH_{c2}/dT \propto T_ch^\prime  \to 0$.

	 \begin{figure}[h]
 \includegraphics[width=8cm]{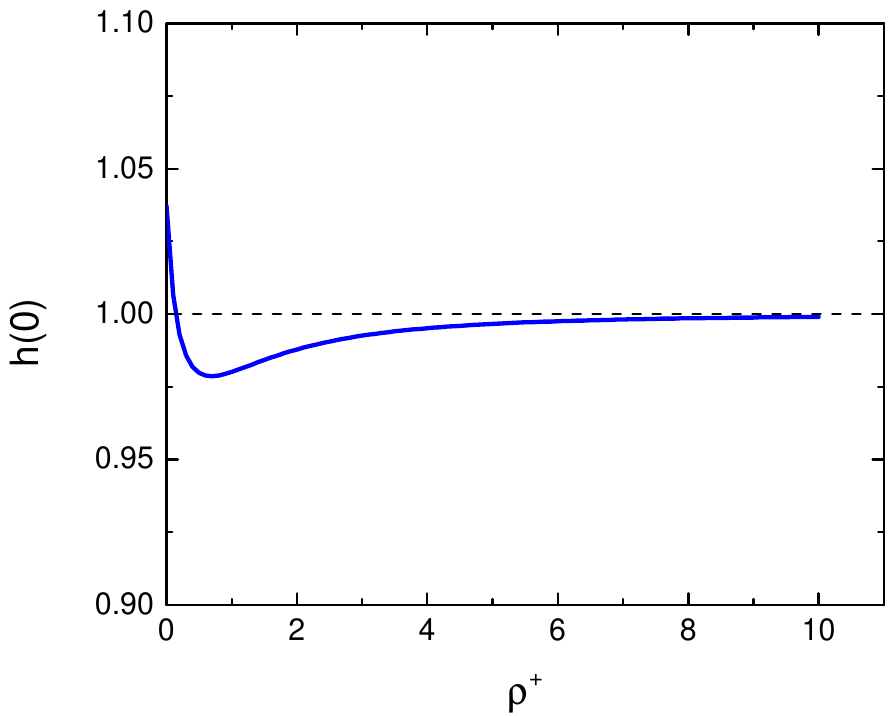}
 \caption{(Color online)   The reduced field $ h(t =0,\rho^+) $ for a d-wave superconductor  calculated with Eq.\,(\ref{eq35}). 
}
  \label{f8d}
 \end{figure}

Next, we calculate the field  at $T=0$. To this end, we transform Eq.\,(\ref{s-c-d}):
\begin{eqnarray}
   -\ln t&=&  \sum_{n=0}^\infty\left(\frac{1}{ n+1/2+\rho^+/2  }- \frac{1}{ n+1/2 }  \right)\nonumber\\
   &+& \sum_{n=0}^\infty \frac{1}{ n+1/2 }-2t \sum_{n=0}^\infty I\,.
\label{eq34}
\end{eqnarray}
The first sum here is expressed in terms of di-gamma functions. 
The divergent sum $\sum(n+1/2)^{-1}$ is truncated   at $n_{max}=\hbar\omega_D/2\pi T$ to give $\ln (2e^{\bm C}\hbar\omega_D/\pi T) $ where $\omega_D$ is the Debye frequency and ${\bm C}=0.577$ is the Euler constant. The last sum in Eq.\,(\ref{eq34}) is replaced with an integral  according to $2\pi T\sum \to \int_0^{\hbar\omega_D} d\hbar \omega $. Since $\gamma=\hbar\omega/\pi T_c +\rho^+$, the integration over $\hbar\omega$ can be replaced with  integration over $\gamma$. Collecting all terms we obtain an equation for $h(0)$ as a function of $\rho^+$:
\begin{eqnarray}
\psi\left(\frac{\rho^++1}{2}\right)+\ln  2  
=\frac{\sqrt{\pi}}{2}\int_0^\infty dz \,{\rm erfc} (z)\ln\left(z^2 h+  \rho_+^2 \right)\,.\nonumber\\   \label{eq35}
\end{eqnarray}
  Fig.\,\ref{f8d} shows that, in fact, $h(0)\approx 1$    for all $ \rho^+ $  within  5\% accuracy.   Physical significance of the shallow minimum in $h(0; \rho^+) $ is not clear. 
  
\begin{figure}[h]
 \includegraphics[width=8cm]{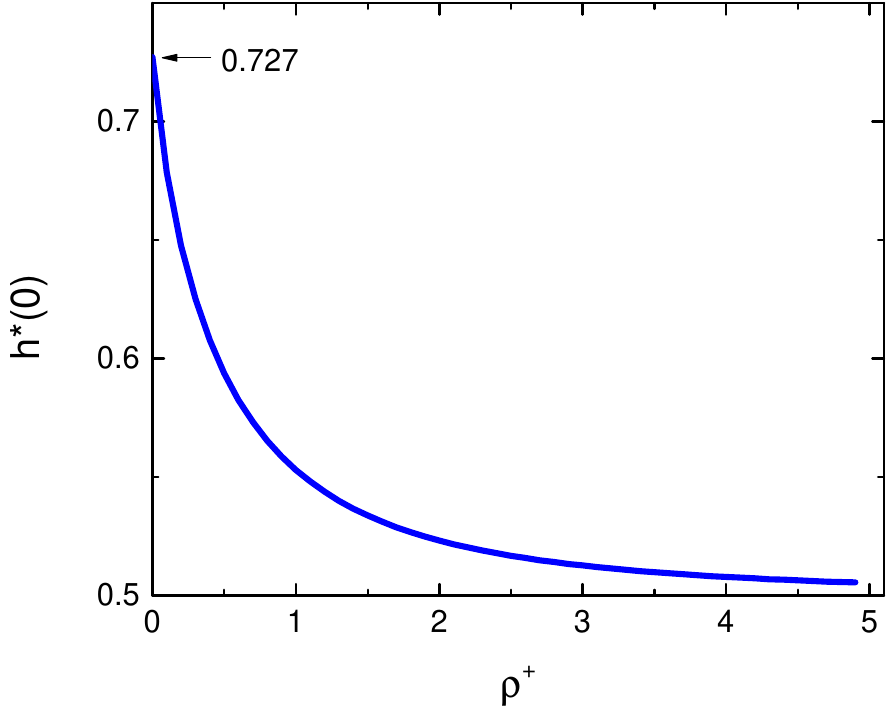}
 \caption{(Color online)   The HW ratio $h^*(0)=h(0)/h^\prime(1)$ as a function of $\rho^+$ for a d-wave superconductor. 
}
  \label{f9d}
 \end{figure}

  Fig.\,\ref{f9d} shows the HW ratio $h^*(0)=h(0)/h^\prime(1)$ as a function of $\rho^+$ for a d-wave superconductor.
We note that the HW ratio in the clean limit at $t=0$ is the same for d- and s-waves (for a Fermi sphere) and for a strong pair breaking it approaches 0.5, as is the case for s-wave.

\section{Discussion} 

To summarize, we have solved the problem of the orbital upper critical field $H_{c2}(T)$ for the isotropic case and any combination of transport and pair-breaking  scattering rates, $\rho$ and  $\rho_m$. The simplicity of the model notwithstanding,   $H_{c2}(T,\rho,\rho_m)$ show a number of interesting features. 

The pair-breaking scattering depresses the slopes of the dimensionless upper critical field $h^\prime$ at $t=1$, just the opposite to what the transport scattering does. The suppression is  pronounced even more in common units since $dH_{c2}/dT\propto  h^\prime(1)T_c$ and $T_c$ is suppressed too. For purely transport scattering, $\rho_m=0$, the slopes increase with increasing $\rho$ as they should. For weak transport scattering, the slopes $h^\prime(1) $ are nearly independent of $ \rho_m$ and for a strong pair breaking (roughly, $ \rho_m>4$) they remain low even if the transport scattering intensifies.

For a strong pair breaking, $h=h(0)(1-t^2)$ with $h(0)$ given in Eq.\,(\ref{h_gapless}) which depends only on the ratio  $\rho/\rho_m$. Then, if in a material the temperature dependence of $H_{c2}$ is close to $(1-t^2)$, one can determine $\rho/\rho_m=\tau_m/\tau$, the ratio of scattering rates, from the experimental $h(0)$. In this case  $ \rho_m\gg 1$ and the transport scattering has practically no effect upon the HW scaled field $h^*(t)=h(t)/h^\prime (1)$

The problem of $H_{c2}(T)$ for the d-wave order parameter in the presence of impurities turns out to be simpler than for s-wave. The reason is that the scattering rates $\rho $ and $\rho_m $ enter the theory only as a sum, see Eq.\,(\ref{Tc0/Tc}) for the $T_c$ suppression and Eqs.\,(\ref{s-c-d}) and (\ref{Igamma})  containing only $\rho^+$.

Intriguing in particular is the similarity of the curves for $H_{c2}(T,\rho,\rho_m,a)$ with account for possible ``freezing out" of the spin-flip scattering by the field, with two-band scenarios without pair-breaking scattering  as discussed, e.g., in Ref.\,\onlinecite{Gur3}. We are far from claiming that our model   can be literally applied to real materials, it is too simple and the field freezing of the pair-breaking is introduced in a profoundly qualitative manner. Still, 
in our view possibility of the field suppression of the spin-flip scattering should not be discarded. In fact, this possibility, if confirmed, makes interpretation of $H_{c2}$ curves even less definite as far as extracting material characteristics from the shape   of these curves.\\

 
This work was supported by the U.S. Department of Energy, Office of Basic Energy Sciences, Division of Materials Sciences and Engineering under contract No. DE-AC02-07CH11358.


\end{document}